\documentclass[12pt]{article}
\usepackage{amssymb} 
\usepackage{epsfig}
\usepackage{float}
\newcommand{\be}{\begin{equation}}
\newcommand{\ee}{\end{equation}}
\newcommand{\bea}{\begin{eqnarray}}
\newcommand{\eea}{\end{eqnarray}}

\begin{document} 

\begin{center}
{\bf THE HISTORY  OF NEUTRINO OSCILLATIONS }\footnote{
A report at the ``Nobel Symposium on Neutrino physics'', Haga Slott, Enkoping, Sweden, 
August 19-24, 2004}

\end{center}
\begin{center}
S. M. Bilenky 
\end{center}
\vspace{0.1cm} 
\begin{center}
{\em Joint Institute
for Nuclear Research, Dubna, R-141980, Russia\\}
\end{center}
\begin{center}
{\em SISSA, via Beirut 2-4, I-31014, Trieste, Italy \\}
\end{center}
      
\begin{abstract}

The early history of neutrino masses, mixing and oscillations 
is briefly reviewed.

\end{abstract}
\section{Introduction}
After many years of heroic efforts of many physicists  we have now
model independent evidence of neutrino oscillations.
The evidence of neutrino oscillations was obtained   
in 
the atmospheric Super Kamiokande experiment \cite{SK}, in the solar SNO 
experiment \cite{SNO} in the reactor KamLAND experiment \cite{Kamland},
and also in 
solar neutrino experiments \cite{Cl,GALLEX,SAGE,SKsol},  atmospheric 
neutrino experiments \cite{Soudan,MACRO}, and
in the first long baseline accelerator K2K experiment \cite{K2K}.

Neutrino oscillations 
are signature of small neutrino masses and 
neutrino mixing.
It took more than 40 years to discover this phenomenon.

The first idea of neutrino oscillations was put forward by B. Pontecorvo in 1957-58
\cite{Pont1,Pont2}.
I worked with B. Pontecorvo 
more than 15 years, starting from the time when majority 
of physicists believed that neutrinos are massless two-component particles.
I will consider mainly the 
evolution of {\em original ideas}
of neutrino masses, mixing and oscillations.

When Pauli introduced neutrino in 1930
he assumed that neutrino ("neutron") 
is a  neutral weakly interacting particle with spin 1/2 and  a mass smaller
than electron mass.
The first method of the measurement of neutrino mass was proposed
in 1933 by Fermi  \cite{Fermi}
and Perrin \cite{Perrin}. They proposed to search for effects 
of neutrino mass via detailed investigation of the high-energy 
part of 
$\beta$-spectra which correspond to the emition of neutrino with a small 
energy.

Usually
effect of the neutrino mass is searched for through the 
investigation of the $\beta$-spectrum of the decay
\be
^{3}\rm{H}\to ^{3}\rm{He}+e^{-}+\bar\nu_{e}
\label{1}
\ee
Up to now no effects of neutrino mass was found in these experiments.
In the first experiments 
for the upper bound of the neutrino mass it was obtained
\cite{numass}

$$m_{\nu}\lesssim 500 \rm{ eV}$$

With further experiments this bound was decreasing and
at the end of the fifties
for the upper bound of the neutrino mass was found the value
\be
m_{\nu}\lesssim (100-200)\,~\rm{eV}. 
\label{-1}
\ee

{\em The two-component neutrino theory}, proposed 
by Landau \cite{Landau}, Lee and Yang \cite{LeeYang} and Salam \cite{Salam}
in 1957 
after the  violation of the parity in the 
$\beta$-decay was discovered \cite{Wu},
{\em was first theoretical  idea about neutrino mass}.

In order to demonstrate the idea of the two-component neutrino
let us consider the Dirac equation for the field of neutrino with mass
$m_{\nu} $
\be
i\,\gamma^{\alpha}\, \partial _{\alpha}\,\nu(x)-m_{\nu}\,\nu(x)=0
\label{0}
\ee

For left-handed and right-handed components 
$\nu_{L}(x)$ and $\nu_{R}(x)$ from Eq.(\ref{0})
we have  two coupled equations
\be
i\,\gamma^{\alpha}\, \partial _{\alpha}\,\nu_{L}(x)-m_{\nu}\,\nu_{R}(x)=0
\label{2}
\ee
and
\be
i\,\gamma^{\alpha}\, \partial _{\alpha}\,\nu_{R}(x)-m_{\nu}\,\nu_{L}(x)=0
\label{3}
\ee
Taking into account the bound (\ref{-1}) it looked natural at the fifties
that neutrino mass is equal to zero.
This assumption was made by Landau, Lee and Yang and Salam.

For $m_{\nu} =0$ from (\ref{2}) and  (\ref{3}) we obtain
 two decoupled Weil equations 
\be
i\,\gamma^{\alpha}\, \partial _{\alpha}\,\nu_{L,R}(x)=0
\label{4}
\ee
and the neutrino field can be in this case
$$\nu_{L}(x)\,~~ \rm{or}\,~~ \nu_{R}(x).$$ 
The was the choice  of the authors of the two-component neutrino theory.

If neutrino field is $\nu_{L}(x)$ ( $\nu_{R}(x)$ )
\begin{enumerate}
\item
The general Hamiltonian of the $\beta$-decay has the form 
\be
{\cal{H}}_{I}^{\beta}=
\sum_{i}G_i \,(\bar{p}\, O_{i}\, n) 
(\bar{e} \,O^{i}\,\frac{1}{2}\,(1\mp\gamma_{5}) \nu)  + 
h.c.,
\label{5}
\ee
where the index $i$ runs over $S,V,T,A,P$ (scalar, vector etc)

Thus, the two-component neutrino theory ensure 
large violation of parity, observed 
in the $\beta$-decay.
\item
Neutrino helicity is equal to -1 (+1) and
antineutrino helicity is equal to +1 (-1) in the case of
$\nu_{L}(x)$ ( $\nu_{R}(x)$ ).
\end{enumerate}

Neutrino helicity was measured in 1958 in a spectacular 
M.Goldhaber et al experiment \cite{GGS}.
In this experiment the circular polarization of $\gamma$-quanta
from the chain of the reactions
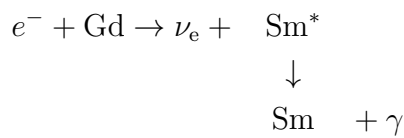
\begin{eqnarray}
e^- + \rm {Gd} \to \nu_e + \null & \rm {Sm}^* &
\nonumber
\\
& \downarrow &
\nonumber 
\\
& \rm{Sm} & \null + \gamma
\nonumber
\end{eqnarray}
was measured. The measurement of the polarization of the
$\gamma$-quanta  allowed 
to determine the longitudinal polarization of neutrino.
It was found  that neutrino is the left-handed particle. 
Thus, the neutrino field is
$\nu_{L}(x)$.

It is interesting to notice that 
equations (\ref{4}) for a massless particle
was discussed by Pauli in his 
encyclopedia article "General Principles of Quantum Mechanics" 
(1933). Pauli wrote that because  equation 
for $\nu_{L}$ ($\nu_{R}$) is not invariant
under space reflection it is "not applicable to the physical reality".

From the point of view of the two-component theory 
large violation of parity in the $\beta$-decay and other 
leptonic processes is ultimately connected with 
equal to zero neutrino mass.  
This point of view changed after 
Feynman 
and Gell-Mann \cite{FeyGel}, Marshak and Sudarshan \cite{MarSud} 
proposed in 1958 $V-A$ theory.

This theory was based on the assumption that
{\em in the Hamiltonian of the weak interaction enter
left-handed components of all fields}. This means
that the
violation of parity in the weak interaction is not connected with 
exceptional properties of neutrinos.
Exist other reasons for left-handed fields in the Hamiltonian.
Moreover,  
after the V-A theory 
it was natural to turn up arguments and consider
neutrino as a particle with 
different from zero mass (see later).

Nevertheless, the two-component neutrino theory 
 was a nice and the simplest theoretical possibility. It
was in a perfect agreement with numerous experiments on the investigation of weak processes. 
From my point of view this was the main reason why
during many years there was a common opinion that 
neutrinos are massless  particles.
The Glashow-Weinberg-Salam Standard Model 
was build under the assumption of massless two-component neutrinos.

\section{B. Pontecorvo}

The first idea of neutrino masses, mixing and oscillations 
was suggested by  B.Pontecorvo in 1957 \cite{Pont1}. He thought that there is 
an 
analogy between 
leptons and hadrons and
he believed that in the lepton world exist phenomenon
analogous  to the famous $K^{0}\rightleftarrows\bar K^{0}$
oscillations.

The only possible candidate were neutrino oscillations.
At that time only one neutrino type was known.  Possible oscillations
in this case are
$$\nu_{L}\rightleftarrows\bar \nu_{L} \,~~\rm{and }\,~~
\bar\nu_{R}\rightleftarrows \nu_{R}$$
According to the two-component neutrino theory the states
$\bar\nu_{L}$ and  $\nu_{R}$
do not exist.
Such states were a problem for B. Pontecorvo.
We will see how he 
solved it.

In 1957-58 R.Davis \cite{Davis}  was doing an experiment  searching 
for production of $^{37} \rm{Ar} $
in the process
\be
\bar \nu_{e} + ^{37}\rm{Cl} \to e^{-} + ^{37}\rm{Ar} 
\label{6}
\ee
with antineutrinos from a reactor.
A rumor reached B.Pontecorvo that Davis observed production 
of $^{37} \rm{Ar}$.
B. Pontecorvo, who was thinking about neutrino oscillations 
at that time, 
decided that production 
of $^{37} \rm{Ar}$  could be due to 
antineutrino$\rightleftarrows$ neutrino transitions in vacuum. 
He published the paper on neutrino oscillations \cite{Pont2}.
In this paper he wrote:
\begin{quote}
"Recently the question was discussed whether there
exist other
{\em mixed} neutral particles beside the $K^0$ mesons, i.e. particles
that differ from the corresponding antiparticles, with the transitions
between particle and antiparticle states not being strictly forbidden.
It was noted that neutrino might be such a mixed particle, and
consequently, there exists the possibility of real neutrino
$\rightleftarrows$ antineutrino transitions in vacuum, provided that
lepton (neutrino) charge is not conserved.
This means that the
neutrino and antineutrino are {\em mixed} particles, i.e., a symmetric and
antisymmetric combination of two truly neutral Majorana particles $\nu_1$
and $\nu_2$."
\end{quote}
 B.Pontecorvo came to the conclusion that 
\begin{quote}
" The flux of neutral leptons 
consisting mainly of antineutrino when emitted from a 
reactor will consist at some distance R from the reactor 
of half neutrinos and half antineutrinos"
\end{quote}
In the fifties Reines and Cowan \cite{ReinesC} 
were doing their famous experiment 
in which $\bar\nu_{e}$ was discovered 
via the observation of $e^{+}$ and neutrons produced in the
reaction
\be
\bar\nu_{e}+p \to e^{+}+n
\label{7}
\ee
In order to see effect of neutrino oscillations 
B. Pontecorvo proposed  that
\begin{quote}
"It will be extremely 
interesting to perform Cowan and Reines experiment at different 
distances from reactor"
\end{quote}

In the paper \cite{Pont2} which  was written 
at the time when the two-component
theory had just appeared and the Davis experiment was not
finished  B. Pontecorvo wrote 
\begin{quote}
"...it is not possible to state apriori 
that some part of the flux can initiate the 
Davis reaction" 
\end{quote}
Thus, he admitted at that time that two-component theory 
can be violated.
Later after the Davis experiment was finished and no production of 
$^37 \rm{Ar}$ was observed B. Pontecorvo  understood that  
due to oscillations neutrino (antineutrino) could transfer 
into $\bar\nu_{L}$ ($\nu_{R}$ ), particles  which do not participate
in the standard weak interaction.
B. Pontecorvo was the first who introduced
the notion of {\em sterile neutrinos} so popular nowadays.

After the second neutrino $\nu_{\mu}$ was discovered 
in the Brookhaven experiment \cite{Brookhaven}
it was very natural (and not difficult ) for 
B. Pontecorvo  to generalize his  idea of neutrino oscillations for the case of 
two neutrinos \cite{Pont3}.
He considered in the paper \cite{Pont3} oscillations into active and sterile states 
$\nu_{\mu}\rightleftarrows
\nu_{e}$, $\nu_{\mu}\rightleftarrows \bar\nu_{\mu L}$ etc.

In the paper \cite{Pont3} B. Pontecorvo 
discussed  oscillations of the solar neutrinos. In 1967
the Davis solar neutrino experiment  only started.
Three years before the first results of  Davis
experiment were published
B. Pontecorvo pointed out that due to neutrino
oscillations the observed flux of solar neutrinos could be two times 
smaller
than the expected flux.
\begin{quote}
"From an observational point of view the ideal object is the sun. If
the oscillation length is smaller than the radius of the sun region
effectively producing neutrinos, (let us say one tenth of the sun radius
$R_\odot$ or 0.1 million km for $^8B$ neutrinos, which will give the
main contribution in the experiments being planned now), direct
oscillations will be smeared out and unobservable. The only effect on
the earth's surface would be that the flux of observable sun neutrinos
must be two times smaller than the total neutrino
flux".
\end{quote}

In the Davis experiment only very rare high energy solar neutrinos mainly 
from the decay  $^{8}B\to^{8}B e^{+}\nu_{e} $ 
can be observed. In \cite{Pont3} B. Pontecorvo wrote
\begin{quote}
"Unfortunately, the relative weight of different thermonuclear reactions in
the sun and its central temperature are not known well enough 
to permit a comparison of the expected and observed
solar neutrino intensities"
\end{quote}
In 1967 it was impossible to envisage 
the NC result of the SNO experiment \cite{SNO} in which the total flux of 
the $^{8}B$ neutrinos was measured. At that time
even neutral current interaction was not known.
It is interesting to notice that in 1988 when 
the SNO experiment was in its initial stage 
B.Pontecorvo enthusiastically supported the experiment.
This  is a letter which B. Pontecorvo wrote at that time.
\begin{quote}
Dr. Walter F. Davidson

High Energy Physics Section,

National Research Council, Canada

Dear Dr. Davidson,

Thank you very much for sending me the  SNO proposal.
Below I am writing a short comment on SNO in the hope 
that opinion of a person who already in 1946 worked in 
Canada on neutrinos may be of some value.
The SNO proposal (1000 tons of $\rm{D}_{2}\rm{O}$ immersed in 
$\rm{H}_{2}\rm{O}$ in a mine 2 km deep) 
in my opinion is a wonderful 
proposal for several reasons.

First it is new in the sense that with the help of 
large $\rm{D}_{2}\rm{O}$ detector immersed in $\rm{H}_{2}\rm{O}$ 
there becomes possible the investigation of reactions
1. $\nu_{e}d\to e^{-}pp$ 2. $\nu_{x}e\to\nu_{x}e $
3. $\nu_{x}d\to \nu_{x}np$ 4.$\bar\nu_{e}d\to e^{+}nn$
5. $\bar\nu_{e}p\to e^{+}n$
with main application to solar and star collapse neutrinos 
(1,2,3) and star collapse antineutrinos 
(4,5).

Second. the proposal is realistic, in a sense that at least one large 
Cerenkov counter filled with $\rm{H}_{2}\rm{O}$ is known to work properly 
(Kamiokande)

Third, the proposal can be realized only in Canada,
where for historical reasons large quantities of
$\rm{D}_{2}\rm{O}$ are available during a period of several years. 

Finally, in my opinion the neutral current reaction 3. 
yielding the total number of neutrinos of all flavors, 
can be investigated in spite of serious difficulties of registration 
of neutrons.

In conclusion the SNO proposal is progressive and 
should be supported by all means. 

Yours sincerely,

Bruno Pontecorvo, 

Dubna August 18, 1988.
\end{quote}

\section{Z.Maki, M. Nakagawa and S. Sakata }

Two-neutrino mixing was proposed in 1962 by
Z. Maki,M.Nakagawa and S. Sakata in a paper \cite{MNS} which 
was practically  unknown for many years.
The 
approach accepted  in this paper was based on the Nagoya model.
According to this model  in hadronic current enter fields of 
three fundamental barions $p$, $n$ and
$\Lambda$. These particles  were considered as bound states of leptons and a boson $B^{+}$
("new sort of matter"):
$$p=<\nu \, B^{+}>,\, n=<e^{-} \, B^{+}>,\, \Lambda=<\mu^{-} \, B^{+}>.$$

A natural consequence of the model was
barion-lepton symmetry, a symmetry of the weak current under the 
exchange
$$\nu \leftrightarrow p,\,~ e^-\leftrightarrow n,\,~ 
\mu^-\leftrightarrow \Lambda. $$

The paper \cite{MNS} was writen at the time when 
there was an indication  \cite{Feinberg},
obtained from the analysis
of the data of experiments on the search for 
$\mu\to e\gamma$,
that
$\nu_{e}$ and  $\nu_{\mu}$ are different particles.
The Brookhaven neutrino experiment \cite{Brookhaven} which proved 
that  $\nu_{e}$ and  $\nu_{\mu}$ are different particles was in preparation at that time.

Possible existence of two different neutrinos was a 
problem for the Nagoya model
(four leptons and three fundamental hadrons).
MNS proposed the following solution of the problem.
The leptonic weak current 
\be
j_{\alpha}=2\,(\bar\nu_{e L}\,\gamma_{\alpha}\,e_{L} + 
\bar\nu_{\mu L}\,\gamma_{\alpha}\,\mu_{L})
\label{8}
\ee
determines weak neutrinos $\nu_e$ and $\nu_{\mu}$.
They wrote
\begin{quote}
"The definition of the particle state of neutrino is quite arbitrary; 
we can speak of "neutrinos" which are different of weak neutrinos 
but expressed
by the linear combinations of the latter. 
We assume that there exist a representation which defines 
{\em the 
true neutrinos} $\nu_1$ and
$\nu_2$ through orthogonal transformation:
\bea
\nu_1 = +\nu_e \cos \delta + \nu_\mu \sin \delta
\nonumber\\
\nu_2 =  -\nu_e \sin \delta + \nu_\mu \cos \delta 
\label{9}
\eea
The true neutrinos should be so defined that $B^{+}$
can be bound to $\nu_1$ but can not be bound to 
$\nu_2$ "
\end{quote}

Thus, MNS proposed  modified  Nagoya model:

$$p=<\nu_{1}\, B^{+}>,\,~ n=<e^{-} \, B^{+}>,\,~
\Lambda=<\mu^{-} \, B^{+}>$$

To the lepton current, written in terms of ``true neutrinos'',
\be
j_{\alpha}=2\,(\bar\nu_{1L}\,\gamma_{\alpha}\,e_{L}\,\cos \delta  + 
\bar\nu_{1L}\,\gamma_{\alpha}\,\mu_{L}\, \sin \delta) +....       
\label{10}
\ee
corresponds the hadronic current
\be
j_{\alpha}=2\,(\bar p_{L}\,\gamma_{\alpha}\,n_{L}\,\cos \delta  + 
\bar p_{L}\,\gamma_{\alpha}\,\Lambda_{L}\, \sin \delta )+....       
\label{11}
\ee
which was identical  to Gell-Mann-Levy current 
\cite{GellLevi}. In fact this was a  pre-Cabibbo discussion of
the hadron mixing.

Further, MNS assumed that there exist an additional interaction 
of $\nu_{2}$
with a field $X$ of heavy bosons
\begin{equation}
{\cal L} = g \,\overline{\nu}_2
\nu_2 \, X^{+} X
\label{12}
\end{equation}
which provide difference of the masses of $\nu_{2}$ and  
$\nu_{1}$.

In connection with the Brookhaven neutrino experiment
MNS wrote
\begin{quote}

"Weak neutrinos \footnote{Let us stress that according to MNS $\delta$ is
the Cabibbo angle.}
\bea
\nu_e = \nu_1 \cos \delta - \nu_2 \sin \delta\nonumber\\
\nu_{\mu} =  \nu_1 \sin \delta + \nu_2 \cos \delta 
\label{13}
\eea
are {\em not stable} due to occurrence of virtual transition
$\nu_{e}\leftrightarrow \nu_{\mu}$ caused by the interaction 
(\ref{12}).
A chain of reactions
\bea
\pi^{+} \to \mu^{+} +\nu_\mu \nonumber\\
\nu_\mu + Z \to Z^\prime +
      \left( \mu^- \rm{and/or}\,~ e^- \right)
\nonumber
\eea
is useful to check the two-neutrino hypothesis only when
$|m_{\nu_2} -
m_{\nu_1}| \leq 10^{-6}$ MeV under the conventional geometry of the
experiments .
Conversely, the absence
of $e^-$ will be able
not only to verify two-neutrino hypothesis but also to provide an upper
limit of the mass of the second neutrino $\nu_2$ if the present scheme
should be accepted"
\end{quote}
\section{B. Pontecorvo and collaborators}
The first phenomenological theory of two-neutrino mixing
was proposed by
V. Gribov and B. Pontecorvo in 1969 \cite{GribovP}.
They assumed that the left-handed fields
$\nu_{e L}$ and $\nu_{\mu L}$
enter not only into the weak interaction but also into a mass term
(which they called an additional neutrino interaction).

Let us notice that there was a wide-spread prejudice at that time 
that in the case 
of left-handed neutrino fields 
neutrino masses must be equal to zero. 
This is correct if total lepton
number is conserved. In \cite{GribovP}  it was shown
that 
if total lepton number is changed by 2
neutrino masses can be introduced even in the case of left-handed 
fields in the Lagrangian.

Gribov and Ponecorvo assumed that the following neutrino mass term  
enters into the Lagrangian
\begin{eqnarray}
{\cal L^{\rm{M}}} = -\frac{1}{2}( m_{e\bar e}\,\overline{(\nu_{e L})^{c}}\,
\nu_{e L}+ 
m_{\mu\bar \mu}\,\overline{(\nu_{\mu L})^{c}}\,\nu_{\mu L}+ \nonumber\\
m_{e\bar \mu}\,(\overline{(\nu_{e L})^{c}}\,\nu_{\mu L}+ 
\overline( {\nu_{\mu L})^{c}}\,\nu_{e L} )) +\rm{h.c.}
\label{14}
\end{eqnarray}
Here
$m_{\mu\bar \mu},\,~m_{e\bar e}$ and $m_{e\bar \mu}$
are  real parameters and $(\nu_{l L})^{c}=C\,(\bar\nu_{l L})^{T}$ is 
the conjugated field.

After the diagonalization of the mass term (\ref{14}) 
the following mixing relations were obtained
\bea
\nu_{e L} =\cos \theta\,~ \nu_{1L} + \sin \theta\,~ \nu_{2L}\nonumber\\
\nu_{\mu L} =-\sin \theta\,~ \nu_{1L} + \cos \theta\,~ \nu_{2L}.
\label{15}
\eea
Here
$\nu_{1,2}$ are fields of {\em Majorana neutrinos} 
with masses $m_{1,2}$. 
The mass term (\ref{14}) is called the Majorana mass term.

Possible oscillations in the case of the mixing (\ref{15}) 
are $\nu_{e}\rightleftarrows\nu_{\mu}$ and  
$\bar\nu_{e}\rightleftarrows\bar\nu_{\mu}$. There are
no transitions into sterile 
states in the case of the Majorana mass term.
This was one of the main idea of Gribov and Pontecorvo.

It was shown in \cite{GribovP} that the observables 
(the mixing angle $\theta $ and Majorana neutrino masses $m_{1,2}$) are 
connected with the parameters of the theory
by the relations
\be
\tan 2\theta = \frac{2\,m_{e \bar\mu}}
{m_{\mu \bar\mu} - m_{e \bar e}},
\label{16}
\ee

\be
m_{1,2}=\frac{1}{2}\,
\left |m_{\mu\bar \mu}+m_{e\bar e}\mp \sqrt{(m_{\mu\bar \mu}-m_{e\bar e})^{2} +
4\,m^{2}_{e\bar \mu}}\right|.
\label{17}
\ee
From (\ref{16}) it follows that
if $\mu-e$ symmetry 
\be
m_{\mu\bar \mu}=m_{e\bar e};\,~~m_{e\bar \mu}\not=0
\label{18}
\ee
holds, $\theta=\pi/4$ (maximal mixing).
Vacuum oscillations of solar neutrinos 
were discussed in \cite{GribovP}   
in this case.

The full phenomenological theory of neutrino mixing and the theory
of neutrino oscillations in vacuum were developed in the seventies.

In the papers \cite{BilPont1}
neutrino mixing was introduced in 
analogy with Cabibbo-GIM mixing of quarks (lepton-quark analogy).

The main idea of the paper \cite{BilPont1} was the following: neutrinos 
like all other 
fundamental fermions (leptons and quarks) are massive particles.
A mixing of massive fermions is a general feature of gauge theories 
with spontaneous violation of
symmetry. Thus,  it is natural
to assume that 
{\em phenomenon of mixing is common for quarks and massive neutrinos}.
For the case of two neutrinos we have
\bea
\nu_{e L} &=&\cos \theta\,~ \nu_{1L} + \sin \theta\,~ \nu_{2L}\nonumber\\
\nu_{\mu L} &=&-\sin \theta\,~ \nu_{1L} + \cos \theta\,~ \nu_{2L}
\label{19}
\eea
Here
$\nu_{1,2}$ are 4-component 
fields of {\em Dirac neutrinos} with masses $m_{1,2}$.

Possible values of the neutrino mixing angle
were discussed in \cite{BilPont1}. We have concluded
\begin{quote}
"... it seems to us that the special values of the mixing 
angle $\theta=0$ (the usual scheme in which muonic charge is strictly
conserved) and $\theta=\pi/4$ (maximal mixing) are of 
the greatest interest."
\end{quote}
The Dirac mass term has the form
\be
\mathcal{L}^{\mathrm{D}}=- \sum_{l l'}\bar \nu_{l'R}\,M_{l'l}^{\mathrm{D}}
\,\nu_{lL} +\mathrm{h.c.}
\label{20}
\ee
where $M_{l'l}^{\mathrm{D}}$ is a complex matrix.
If the mass term (\ref{20}) enter into Lagrangian, 
the total lepton number $L $ is conserved. 
Due to the conservation of 
$L$
in the case of the Dirac mass term there are no transitions into sterile states. 
For two neutrinos $\nu_{e}$ and $\nu_{\mu}$ the possible 
oscillations are 
$\nu_{e}\rightleftarrows \nu_{\mu}$ and
$\bar\nu_{e}\rightleftarrows \bar\nu_{\mu}$,
the same as in the Majorana case.

In \cite{BilPont2} we have introduced the most general mass term.
If we assume that 
$\nu_{l L}$ and $\nu_{lR}$ enter into the mass term 
and the total lepton number $L$ is 
not conserved for the mass term we obtain
\begin{eqnarray}
\mathcal{L}^{\mathrm{D+M}}=
-\frac{1}{2}\,\sum_{l,l'}( \overline{\nu_{l'L})^{c}}\,M^{\mathrm{L}}_{l'l}
 \nu_{lL}-  \sum_{l'l'}\bar \nu_{l'R}\,M_{l'l}^{\mathrm{D}}\, \nu_{lL}
 \nonumber\\
-\frac{1}{2}\,\sum_{l,l'}\bar\nu_{l'R}\,M^{\mathrm{R}}_{l'l}
(\nu_{lR})^{c} +\mathrm{h.c.}  
\label{21}
\end{eqnarray}
Here
$M^{\mathrm{L}}$ and $M^{\mathrm{R}}$ are complex symmetrical matrices 
and $M^{\mathrm{D}}$ is a complex matrix. The Eq.(\ref{21})
includes left-handed Majorana mass term, the Dirac mass term and the right-handed 
Majorana mass term. It 
is called the Dirac and Majorana mass term.

For the mixing in the general case of $n$ flavors
from (\ref{21}) we have
\begin{eqnarray}
\nu_{lL} &= &\sum_{i=1}^{2n} U_{li} \nu_{iL}\nonumber\\
(\nu_{lR})^{c} &=& \sum_{i=1}^{2n} U_{\bar l i} \nu_{iL},
\label{22}
\end{eqnarray}
where $U$ is an unitary 2n$\times$2n mixing matrix, 
$\nu_{i}$ is the field of the Majorana neutrino with mass $m_{i}$.
Now after the LEP experiments 
we know that $n=3$.

From (\ref{22}) it follows that in the case of small $m_{i}$ 
transitions $\nu_{l}\to \nu_{l'}$ (flavor-flavor) 
and 
$\nu_{l}\to \bar\nu_{l'L}$ (flavor -sterile)  
are possible. Vacuum oscillations of the solar neutrinos in the general 
case of the Dirac and Majorana mass term were
considered in \cite{BilPont2}.
As it is  well known the Dirac and Majorana mass term 
is the framework for the see-saw mechanism
of neutrino mass generation.

In the seventies it was developed also the theory of neutrino oscillations 
in vacuum which is widely used
today for analysis of the data of neutrino oscillation experiments.

In the case of neutrino 
mixing the lepton numbers $L_{e}$, $L_{\mu}$ and $L_{\tau}$ are
not conserved. 
What are flavor neutrinos $\nu_{e}$, $\nu_{\mu}$, $\nu_{\tau}$ and
corresponding antineutrinos in this case?
From the very beginning we 
determine
flavor neutrinos and antineutrinos as particles 
which take part in the
standard CC weak processes with corresponding leptons. 
For example,  neutrino which is produced
together with $\mu^{+}$ in the decay $\pi^{+}\to\mu^{+}+\nu_{\mu}$  
is muon neutrino $\nu_{\mu}$,
electron antineutrino $\bar \nu_{e}$ produces
$e^{+}$ in the process $\bar \nu_{e}+ p \to e^{+}+n $ 
etc. 

The states of flavor neutrinos are given by
\be
|\nu_l\rangle
=
\sum_{i} U_{li}^* \,~ |\nu_i\rangle,
\label{23}
\ee
where $|\nu_i\rangle$ is the state of neutrino with mass $m_{i}$,
momentum $\vec{p}$ and 
energy
$E_{i}= \sqrt{p^{2}+ m^{2}_{i}}\simeq p +\frac{m^{2}_{i}}{2\,p}$.
Thus, flavor neutrinos are described by {\em mixed  coherent states}.

The relation (\ref{23}) is based on the assumption that
neutrino mass-squared differences are so small that
due to the uncertainty relation it is 
impossible to distinguish production (detection)
of neutrinos with different masses.
This condition can be presented in the form
\be
L_{osc}\gg d,
\label{24}
\ee
where d is a quantum mechanical dimension of a neutrino source
and 
\be
L_{osc}=4\,\pi\,\frac{E}{\Delta m^{2}}
\label{25}
\ee
is the oscillation length.
If this condition is satisfied, neutrino cross sections and decay probabilities 
are given by the Standard Model.

If we apply now to the flavor states the evolution equation 
of the field theory
\be
i\frac{\partial |\Psi(t)\rangle} {\partial t}={\mathrm H}\,
|\Psi(t)\rangle
\label{26}
\ee
we come to the standard expression for the transition probability
\be
{\mathrm P}(\nu_{l}\to \nu_{l'}) =
|\delta_{ll'} +\sum_{i\geq 2} U_{l' i}  U_{l i}^*
\,~ (e^{- i \Delta m^2_{i 1} \frac {L} {2E}} -1)|^2 .
\label{27}
\ee
Here $L$ is the source-detector distance, $E$
is neutrino energy and $\Delta m^2_{i 1} =  m^2_{i}- m^2_{1}$.

Necessary condition for the observation of neutrino oscillations
has the form 
\be
\Delta m^2_{i 1} \,\frac {L} {E}\ge 1
\label{28}
\ee
From this condition we see that 
experiments on the search for neutrino
oscillations have enormous sensitivity 
to neutrino mass squared differences (for example, reactor experiments of the KamLAND type 
with $L\simeq 100$ km and $E \simeq 1$ MeV are sensitive to 
$\Delta m^2 \simeq 10^{-5}\rm{eV}^{2}$ etc)

For us this was the main reason and motivation
for neutrino oscillations experiments:
due to interference nature of the phenomenon of 
neutrino oscillations and
possibility to perform neutrino experiments with  large 
values of the parameter $\frac {L} {E}$
investigation of neutrino oscillations is extremely 
sensitive method to search for small $\Delta m^2$.
This strategy brought success. 
We summarized it in the first review on neutrino oscillations
published in 1977 \cite{BilPont3}.
Except papers of B.Pontecorvo, MNS and B.Pontecorvo and collaborators 
at that time it was published only a few papers on neutrino oscillations (\cite{BahFrau,FritMin,
ElizSwif, BilPont4} ).

At the end of the seventies a common interest to the problem of 
the neutrino mass and mixing started.
It was connected with the appearance of the GUT models and the invention 
of the see-saw mechanism of the neutrino mass generation.
Neutrino masses started to be considered as a signature of 
a new, beyond the Standard Model physics. 
At that time special reactor and accelerator neutrino 
experiments on the  search for neutrino oscillations started.
The investigation of matter effects in the case of neutrino mixing in the seventies 
\cite{Wolf} and 
disclosure of the MSW effect in the eighties
\cite{MikhS} had very important impact on the field.

\section{Conclusion}

Today all existing neutrino oscillation data with the 
the exception of the data of LSND experiment \cite{LSND}, 
which need confirmation, are described by the three-neutrino mixing.

For neutrino oscillation parameters
the following values were obtained \cite{SK,Kamland,CHOOZ}:
\bea
\Delta m^2_{2 1}=(8.2 ^{+0.6}_{-0.5})\cdot 10^{-5}\rm{ eV}^{2};\,~
\tan^{2}\theta_{12}=(0.40 ^{+0.09}_{-0.07})\nonumber\\
1.9\cdot 10^{-3}\leq\Delta m^2_{3 2}\leq 3.0\cdot 10^{-3} \rm{ eV}^{2};\,~
\sin^{2}2\,\theta_{23}\geq 0.90\nonumber\\
\sin^{2}\theta_{13}\leq 5\cdot 10^{-2}\,~~~~~~~~~~~~~~~~~~
\label{29}
\eea

Thus, neutrino oscillation parameters satisfy inequalities
\be
\Delta m^2_{2 1}\ll\Delta m^2_{32};\,~~
\sin^{2}\theta_{13}\ll 1
\label{30}
\ee
It follows from (\ref{30}) (see \cite{BGG}) that
the dominant transitions, governed by $\Delta m^2_{3 2}$,
are $\nu_{\mu}\to \nu_{\tau}$ and $\bar\nu_{\mu}\to \bar\nu_{\tau}$.
Dominant transitions, governed by $\Delta m^2_{2 1}$,
 are $\nu_{e}\to \nu_{\mu,\tau}$ and  $\bar\nu_{e}\to \bar\nu_{\mu,\tau}$.
This is a present-day picture of neutrino oscillations.

The main question to be answered what
physics was discovered? What are implications of the discovered phenomenon?
From my point of view to answer these fundamental questions 
we need to know

\begin{enumerate}
\item
Are massive neutrinos Majorana or Dirac particles?
\item

What is the neutrino mass spectrum 
(hierarchical,  inverted,  degenerate etc)?
\item
What is the mass of the lightest neutrino ?

\item
How many massive neutrinos exist in nature? Are sterile neutrinos
exist?

\item
What is the value of the parameter $\sin^{2}\theta_{13}$?

\item
What is the value of CP phase?

\item
What are precise values of oscillation parameters?
\item
...
\end{enumerate}

I think that  the first period of the investigation of neutrino properties is
basically finished.  In spite it took many years to discover neutrino oscillations,
in a sense up to now we have been lucky.
Seems that the next decisive step will be difficult and will require a lot of efforts.  
Existing data allows us to conclude that 
to probe the nature of neutrinos via the investigation of neutrinoless double
$\beta$-decay is a challenge. It will be extremely difficult to observe 
this process
if Majorana neutrino mass spectrum is hierarchical,
which is a plausible possibility. The solution of other problems also look
as a difficult and challenging task. 
After the discovery of neutrino masses and mixing we know, however, 
what to look for.
This is a great advantage of the present
stage.

The history of neutrino oscillations is an illustration of
a complicated and thorny way of science. 
\begin{itemize}
\item Correct pioneer ideas sometimes can
have wrong basis or can be accompanied by wrong one.
\item
Courageous general ideas 
 have good 
chances to be correct.
\item
Analogy is an important guiding principle in physics.
\end{itemize}

I will finish with the citation from the S.L. Glashow report \cite{Glashow} 
at the Venice "Neutrino Telescope Workshop" ( March 2003). 
\begin{quote}
"...If only Bruno Pontecorvo could have seen how far we have come 
towards understanding the pattern of neutrino masses and mixing! Way 
back in 1963 he was among the first have envisaged the possibility of
neutrino flavor oscillations. For that reason. the analog to the 
Cabibbo-Kobayashi-Maskawa matrix pertinent to neutrino 
oscillations should be known as the PMNS matrix to honor 
four neutrino visionaries: Pontecorvo, Maki, Nakagawa, and Sakata"
\end{quote}

\end {document}